\documentclass[12pt]{article} 

\usepackage{mathbbol,amssymb,latexsym,amsfonts,amsmath,amsthm}

\usepackage{amsfonts}
 \usepackage{amssymb}
 \usepackage{amsmath}

\def\pmx{\begin{pmatrix}}
\def\emx{\end{pmatrix}}
\def\bsq{\begin{subequations}}
\def\esq{\end{subequations}}
\def\be{\begin{eqnarray}}
\def\ee{\end{eqnarray}}
\def\bee{\begin{eqnarray*}}
\def\eee{\end{eqnarray*}}
\def\bal{\begin{align}}
\def\eal{\end{align}}

\newcommand{\mfr}[2]{{\textstyle\frac{#1}{#2}}}

\newtheorem{thm}{Theorem}
\newtheorem{lemma}[thm]{Lemma}

\def\bra{\langle}
\def\ket{\rangle}

\def\kb{ \ket \bra }

\def\half{{\textstyle \frac{1}{2}}}

\def\mm{ \! - \! }

\def\nn{\nonumber}

\def\s2{\tfrac{s}{2}}

\def\mm{ \! - \! }

\def\qed{\qquad{\bf QED}}

\newcommand{\proj}[1]{ | #1 \kb  #1|}

\title{{\bf \large Improved Gap Estimates for  Simulating \\ Quantum Circuits  by 
Adiabatic Evolution\thanks{$\copyright $ by authors.   Reproduction permitted for
non-commercial purposes.}}}

\author{Percy Deift\thanks{Partially supported   
   by the National Science  Foundation under Grants  DMS-0296084 and  DMS-0500923} \\{\small Courant Institute of Mathematical Sciences,
251 Mercer St.,   New York, NY 10012} \\ {\small  deift@courant.nyu.edu}
 \and Mary Beth Ruskai  \thanks{Partially supported   
   by the National Science  Foundation under Grant  DMS-0314228 
and by  the National Security Agency  and
 Advanced Research and Development Activity   under
Army Research Office   contract number    DAAD19-02-1-0065.} 
  \\  {\small Department of Mathematics,
     Tufts University,
       Medford, MA 02155} \\
     {\small     Marybeth.Ruskai@tufts.edu}
     \and Wolfgang Spitzer
 \\{\small    Department of Physics, International University of Bremen} \\
  {\small Campus Ring 8, 28759 Bremen, Germany} \\
{\small w.spitzer@iu-bremen.de}}

\date{\small \today}
\begin{document}

\maketitle   
 
\begin{abstract}
We use elementary variational arguments    to prove,
and improve on,   gap estimates  which arise in simulating 
quantum circuits by adiabatic evolution.
\end{abstract}
 
There are several models for quantum computation \cite{NC}.
The quantum Turing machine model
and the quantum circuit model, are equivalent in the sense that
any algorithm that runs in polynomial time in one  
requires only polynomial time in the other.   There are also several
``one-way''   measurement-based  models \cite{J},  such as 
  the cluster state model \cite{RBB},  which can simulate any 
polynomial time quantum circuit in polynomial time.   

In \cite{FGGS}, Farhi, et al
introduced quantum computation by adiabatic evolution of a
Hamiltonian and showed that it could be simulated {\em by} a suitable quantum 
circuit.    In this model, the time required is assumed to depend inversely
on the square of the eigenvalue gap.\footnote{This assumption does not take into
account the fact that higher order terms in the asymptotic expansion may be
needed, or their possible growth as the Hamiltonian changes with the size
of the problem \cite{SMS}.   Although this may affect the time estimates, it seems unlikely
to do more than change the order of the polynomial.   In this note, we deal only
with gap estimates and not with time estimates. 
For further details and
references on issues involved see the report of the workshop at
http://www.perimeterinstitute.ca/activities/scientific/PI-WORK-6/related\_links.php}
 In \cite{ADKLLR}, a method was given for simulating an arbitrary quantum circuit with
 $L$ gates by the  adiabatic evolution of a Hamiltonian in time which is polynomial in $L$,
 using a simple ``clock'' model to construct the Hamiltonian.   
  Some modifications \cite{KKR,Siu} of the Hamiltonian construction
  have been considered without introducing  different techniques for estimating 
  the gap.
   
 In \cite{ADKLLR}, the techniques used to prove the gap estimates are rather 
 complicated.  In this
 note we show that very elementary techniques suffice, and that  one of the bounds
 can be improved by a factor of $L$.   To make precise statements, we need some
 notation.
 
Let  $| e_k \ket$  denote  the standard basis for ${\bf C}^d$;
in particular, $|e_1\ket = (1,0, \ldots 0)^T$.
Let  $-\Delta_d$ denote the discrete Laplacian with Neumann boundary conditions, i.e.,
 \be  \label{delta}
       -\Delta_d =   \left[\begin{array}{rrrrrr} \mfr12 &-\mfr12&0&0&\cdots &0\\[1ex]
                                       -\mfr12&1&-\mfr12&0&\cdots& 0\\[1ex]
				       0&-\mfr12&1&-\mfr12&0&\vdots\\[2ex]
				       \vdots&\ddots&\ddots&\ddots&\ddots&\vdots\\[2ex]
				       0&\cdots&0&-\mfr12&1&-\mfr12\\[1ex]
				       0&\cdots& 0&0&-\mfr12&\mfr12
				       \end{array}\right] \, . 
\ee
The Hamiltonian used in Lemma 3.5 of  \cite{ADKLLR} can be written as
\be  \label{Hs}
      H_0(s) = s (-\Delta_d) + (1-s) I_d - (1-s) \proj{e_1}
\ee
and the block  diagonal Hamiltonian in Lemma 3.12  as
\be
    {\mathbb H}(s) = \bigoplus_{j = 0}^{2^m-1} H_j(s),
\ee
where  
\be   \label{H2s}
      H_j(s) = s (-\Delta_d) + (1-s) I_d  + [b_j - (1-s)] \proj{e_1}.
\ee
with $b_j$ an integer $ \geq 1$ for $j \geq 1$, and $b_0 \equiv 0$.
Let  $ \lambda_1(s) <  \lambda_2(s)  \ldots$ denote the eigenvalues
of $H_0(s) $ and $\Lambda_k(s) $ the eigenvalues of 
$ {\mathbb H}(s) $ also in increasing order.   One is
interested in  $g(s) = \lambda_2(s) - \lambda_1(s) $
and $G(s) = \Lambda_2(s) - \Lambda_1(s) $,
the energy gaps between the two lowest states of  these Hamiltonians.   

The eigenvalue equation  $H_j |u \ket = \lambda |u \ket $  written in terms
of the vector components $u_k$ of   $ |u \ket $ is
\be   \label{eq:eval}
       \s2 \, (u_{k-1} + u_{k+1} )  = (1 - \lambda) \, u_k   \qquad   k = 2, 3, \ldots d \mm 1
\ee
 subject to the boundary conditions
 \be   \label{eq:bc}
      u_1 (b_j +  \s2- \lambda) = \s2 \, u_2   \qquad   
      (1 -  \s2 - \lambda) u_d = \s2 \, u_{d-1} .
 \ee
 This is a second order difference equation with constant coefficients
 subject to boundary  conditions. 
It  can be solved exactly by elementary methods entirely analogous to those used
 to solve the ``particle in a box''  boundary value problem.
For  general $s, b_j$ the algebra can become somewhat tedious and we need 
 only bounds on the lowest eigenvalues, not the full spectrum.    Good estimates
 on the gaps can be obtained from one special case and a simple variational
 argument.
 
\begin{lemma}  \label{lemma}
The lowest eigenvalue $\mu_0$ of 
 $ -\Delta_{d} + \half  \proj{e_1}$ satisfies
 $1 > \mu_0 > \frac{1}{d^2}$.
\end{lemma}

\begin{thm}  \label{thm1}
The energy gaps   for the Hamiltonians given by \eqref{Hs} and \eqref{H2s} 
are both $O(d^{-2})$; in fact,   $g(s) > 1/2d^2$, and $G(s) > 1/2d^2$.
\end{thm}
When $d = L+1$,  part (a) is equivalent to Lemma~3.5 of  \cite{ADKLLR} 
and part (b) improves   Lemma~3.12 of  \cite{ADKLLR} by a factor of $L$.  
Since we are interested in large $d$, we will henceforth not distinguish
between estimates involving $d$ and $d \pm 1$.

\noindent{\bf Proof of Theorem~\ref{thm1}:}  
Consider two simple trial functions using the ground states of $H_0$ at the endpoints
$s = 0,1$.  First,
\be  \label{v1}
 \bra e_1, H_0(s) \, e_1 \ket =  s  \bra e_1, (- \Delta_d )  \, e_1 \ket = \half s.
 \ee
Let $ |v \ket$ denote the normalized constant vector with elements $v_k = \frac{1}{\sqrt{d}} $.
Then
\be   \label{v2}
    \bra v ,  H_0(s)   v \ket  =  (1-s) \tfrac{d-1}{d} = (1-s) - \tfrac{1}{d} (1-s).
\ee
Thus, $\lambda_1(s) \ \leq \min\{ \half s, (1-s) \frac{d-1}{d} \}$ and the two curves cross
at  $s_c$ where $\half < s_c = \frac{2d-2}{3d-2} < \frac{2}{3}$ for $d > 2$.
Next, note that by the max-min principle
\be  \label{lam1}
    \lambda_2(s) \geq \inf_{u \, \in \, e_1^{\perp} }\bra u, H_0(s) \, u \ket  =  (1-s) + s \mu_0.
\ee
Since $\mu_0 < \tfrac{3}{2}$,  one finds that  $ \half(2-3s) + s \mu_0$
is decreasing on $[0,s_c]$ and attains its minimum at $s_c$.  Thus
\be   \label{varbd}
     g(s) & = &   \lambda_2(s) - \lambda_1(s)   
              \geq  ~ \begin{cases}  \half(2-3s) + s \mu_0    & s  \leq s_c \\
              \tfrac{1}{d} (1-s) + s \mu_0 &  s > s_c
           \end{cases}   \nn \\
             & \geq & s_c \mu_0 \geq   \frac{1}{2d^2}    .
     \ee
This proves part (a).   To prove part (b) observe that the condition $b_j \geq 1$
 implies that for $j \geq 1$ the Hamiltonian \eqref{H2s} satisfies
\be  \label{Hjest}
   H_j(s) & \geq & s (-\Delta_d) + (1-s) I_d +s \proj{e_1}  \nn \\
       & \geq & s \big(-\Delta_d +  \half \proj{e_1} \big) + (1-s) I_d  \nn  \\
      &  \geq & s \mu_0 + (1-s).
       \ee
 Now, $\Lambda_2(s)$ is the minimum of $\lambda_2(s) $ and 
 the lowest eigenvalue of $H_j(s)$ with  $j \geq 1$.
Therefore, \eqref{lam1} and  \eqref{Hjest} imply that
$\Lambda_2(s) \geq s \mu_0 + (1-s)$ as well.   Since $\Lambda_1(s) = \lambda_1(s)$,
the argument above implies that gap for $ {\mathbb H}(s) $ satisfies
$G(s) \geq  1/2d^2  $.  
 
\bigskip 
\noindent{\bf Proof of Lemma~\ref{lemma}:}   The eigenvalue problem for
  $ -\Delta_{d} + \half   \proj{e_1}$ is equivalent to setting $s = 1$, and $b = \half$
  in \eqref{eq:eval} and \eqref{eq:bc}.
 We first look for  solutions of the form  $u_k = e^{ik \theta} - e^{-ik \theta}  $ with
 $0 < \lambda < 2$.    When $\lambda = 1 - \cos \theta$, \eqref{eq:eval} 
and  the first boundary condition are satisified.  One
can verify that for $\theta = \frac{2n-1}{2d+1} \pi $, the second condition in \eqref{eq:bc}
holds  for
$n = 1,2, \ldots, d$.   Since this gives $d$ linearly independent solutions in
the range  $0 < \lambda < 2$,  the $d$  eigenvalues  are
$1 - \cos  \big( \frac{2n-1}{2d+1} \big) \pi$ for $n = 1,2, \ldots, d$.    
The smallest eigenvalue is 
\be
\mu_0 = 1 - \cos  \big( \tfrac{1}{2d+1} \big) \pi =  \frac{\pi^2}{8d^2} - O\big( \frac{1}{d^4}\big)
     > \frac{1}{d^2}.
\ee
for $d$ sufficiently large.   \qed

As a final remark, we note that 
  $-\Delta_d   = \half X^\dag X$   with  
 $ X =   \left(\begin{array}{rrrrr} 1 & - 1 & 0 & \cdots & 0 \\
				            0 & 1 & -1 & \cdots & 0 \\
				          \vdots & \ddots  &  \ddots & \ddots & \vdots \\
				           0  & \cdots & 0 & 1 & -1  \end{array}\right)$   
so that $X |u \ket$ has elements  $u_k - u_{k+1}$ and
$$\bra u , -\Delta_d \, u \ket =  \half \bra X u, X u \ket = \half \sum_{k=1}^{d-1} |u_k - u_{k+1}|^2. $$
This lends itself to interpreting $ -\Delta_d$ as a lattice analogue of the
kinetic energy, rather than as a $3$-local potential as in  \cite{ADKLLR}.
Moreover, the interpolating Hamiltonians $H_j$ are linear combinations of
$- \Delta$, the Identity $I$, which can shift the spectrum but has no effect
on the gap, and $\proj{e_1}$ whose only effect is to modify the first
boundary condition.     

\bigskip

\noindent{\bf Acknowledgment:}  This work was begun during
 a workshop  at the Perimeter Institute (PI) in  Waterloo, Canada  
 held during MBR's stay at PI.   She is grateful for their support
 and hospitality.

\end{document}